
\magnification=1200
\baselineskip=16 pt
\def\lapprox{\lower 3 pt\hbox{$\buildrel < \over \sim\;$}}
\def\ref{\smallskip\noindent\hangindent=1.0 true cm}
\centerline{\bf A  MODEL FOR THE CHEMICAL EVOLUTION OF THE GALAXY}
\centerline{\bf WITH REFUSES}
\vskip 2 true cm
\centerline{H. J. ROCHA-PINTO$\sp{1,2}$,  L. I. ARANY-PRADO$\sp{1,2}$}
\centerline{W. J. MACIEL$\sp 3$}
\vskip 2 true cm
\centerline{1: Observat\'orio do Valongo, UFRJ, Rio de Janeiro, Brazil}
\bigskip
\centerline{2: Observat\'orio Nacional, CNPq, Rio de Janeiro, Brazil}
\bigskip
\centerline{3: Instituto Astron\^omico e Geof\'\i sico, USP, S\~ao Paulo,
Brazil}
\vskip 5 true cm
Send proofs to:
\bigskip
Dr. Walter J. Maciel
\par
Instituto Astron\^omico e Geof\'\i sico da USP
\par
Av. Miguel Stefano 4200
\par
04301-904 S\~ao Paulo SP
\par
Brazil
\vfill\eject
\noindent{\bf Abstract.} A model is presented for the chemical evolution of
the solar neighbourhood which takes into account three families of
galactic objects, according to their condensation states:
stars, refuses and gas. Stars are defined as every condensed objects with
masses greater than or equal to the minimum mass which ignites hydrogen and
which will give rise to an evolutionary track on the HR diagram to the left
of Hayashi's limit; refuses include the remnants, which are compact objects
resulting from stellar deaths, and the residues, which have masses not large
enough to ignite hydrogen; gas is defined as the mass which can be condensed
to form stars and/or residues. We have developed equations for the mass
evolution of each family, and have studied the gas metallicity distribution
within the framework of the instantaneous recycling approximation, adopting
different initial conditions. In order to constrain the model parameters
we have also used preliminary evaluations of comet cloud masses to investigate
the role of the residues as sinks of heavy elements in the Galaxy.
\vfill\eject
\centerline{\bf 1. Introduction}
\bigskip\noindent
Models for the chemical evolution of galaxies usually include only two
classes of objects, namely {\it stars} and {\it gas}
(cf. Tinsley, 1980). This classification has as main argument the
simplification of the equations, and generally does not cause
significant difficulties in the derivation and interpretation of
the quantities which are effectively compared with the observational data.

On the other hand, more complete formulations already appeared in some of the
early works on chemical evolution (Schmidt, 1959) and also in more
recent treatments of the evolution of our Galaxy (Tinsley, 1981,
Rana and Wilkinson, 1986) and other spiral galaxies (cf. Ferrini et al., 1992).

The main characteristic of these treatments is the inclusion of
non-stellar objects, as our present knowledge makes us sure
that an important quantity of metals is probably
locked up in some galactic objects such as planets, comets, etc. (cf.
Bailey, 1988). Comets are particularly interesting in this respect,
as they seem to be a very common phenomenon associated with star
formation out of a protostellar cloud (Vanysek, 1987a,b). Similar
to the case of interstellar grains, comet formation has probably
a small effect in the mass balance of the Galaxy (cf.
Meusinger, 1992). On the other hand, these
objects  may affect some observational properties such as the
extinction and polarization of visible light (cf. Greenberg, 1974), and
can in principle have an influence on the chemical evolution
of the solar neighbourhood as metal sinks,
as suggested by Tinsley and Cameron (1974),
Van\'ysek (1987a,b), and Stern and Shull (1990).
\par
In the present paper, we have introduced a consistent treatment of
the evolutionary histories of the different families of galactic objects,
taking into account the following condensation states: {\it stars, refuses,}
and {\it gas}. We then study the derived metallicity distribution for
the one-zone model of the solar neighbourhood considering a set of
initial conditions, and compare our results with observed data from
stars. Finally, we make some preliminary calculations of comet
cloud masses, in order to investigate the role of comets as
heavy element sinks in the galactic disk.
\bigskip
\bigskip
\centerline{\bf 2. Basic Equations}
\bigskip
\noindent
The adopted families of stars, refuses and gas are associated with the
following classes of objects, respectively:
\medskip\noindent
{\bf Stars}, which are defined as every condensed object formed with
masses  $m > m\sb l$, where $m\sb l$
is the lower limit for the stellar masses, or
the minimum mass which produces hydrogen ignition and which will give
rise to a track on the HR diagram to the left side of Hayashi's limit.
\medskip\noindent
{\bf Refuses}, which include {\bf remnants}, or compact
objects resulting from stellar deaths, and {\bf residues} of star formation,
which are  objects condensed from the gas, with masses in the interval
$m\sb g < m < m\sb l$, where $m\sb g$
is the maximum assumed mass for the gas (e.g. grains).
\medskip\noindent
{\bf Gas}, which is the mass that can be condensed to form stars
and/or residues.
\bigskip
Adopting a model for the chemical evolution of the
solar neighbourhood with no infall, the total mass of the system is
constant and given by
$$M = M\sb g + M\sb s + M\sb r \eqno(1)$$
where $M\sb g, M\sb s,$ and $M\sb r$  are the
total masses in gaseous, stellar and refuse condensation states,
respectively. The gas  ($\mu$) and refuse ($\kappa$)
mass fractions are defined by
$$\mu \equiv {M\sb g\over M} \eqno(2)$$
and
$$\kappa \equiv {M\sb r\over M} \eqno(3)$$
so that
$$M\sb s = (1 - \mu - \kappa) M. \eqno(4)$$
We will adopt the usual sudden mass loss approximation, where  the stars
undergo the entire process of mass loss after a well-defined lifetime.
Let $w_m$ be the remnant mass
of a star with initial mass  $m$ and lifetime
$\tau\sb m$. The rate of mass locked up in the remnants, due to the death
of the stars which were born at instants given by $(t-\tau\sb m)$
is obtained by
$$L(t)  = \int_{m\sb t}^{m_u} w_m(m) \Psi(t-\tau\sb m)
\Phi(m) dm \eqno(5)$$
and the total ejection  rate due the death of these stars is
$$E(t) = \int_{m\sb t}^{m_u} [m - w_m(m)] \Psi(t - \tau\sb m) \Phi(m)
dm, \eqno(6)$$
where $m\sb t$ is an appropriately chosen turnoff mass;
$m_u$ is the upper limit to the stellar mass, or the
maximum mass admitted to stars;
$\Psi$ is a \underbar{generalized} formation rate, defined as the total
mass condensed into galactic objects per unit time,
and  $\Phi$ is a \underbar{generalized} initial mass function, normalized as
$$\sum^4_{i=1}\int^{x_{i+1}}_{x_i} m \Phi(m) dm = \rho + \gamma +
\zeta + \varepsilon = 1 \eqno(7)$$
where $x_1=0$, $x_2=m\sb g$, $x_3=m\sb l$,
$x_4=m\sb t$, $x_5=m_u$, and
$\rho$, $\gamma$, $\zeta$ and $\varepsilon$ are,
respectively, the first, the second, the third and the fourth
terms in the sum. Each of these four terms,
multiplied by the generalized formation rate  will give respectively:
the formation rate of objects with masses $m<m\sb g$,
considered as gas, $\rho\Psi(t)$; the
formation rate of the residues, $\gamma\Psi(t)$;
and the stellar formation rates,
$\zeta\Psi(t)$ and $\varepsilon\Psi(t)$, relative to stars
with masses in the intervals  $m\sb l<m<m\sb t$ and
$m\sb t<m< m\sb u$, respectively.

It is worth noting that the generalized initial mass function
is not necessarily continuous in the
above intervals. However, we have assumed
its continuity and will take $m\sb g\approx 0$,
so that $\rho\approx 0$. We will assume that every residue is
gravitationally tied with a star and that the formation of stars with
mass $m>m\sb t$ is not accompanied by formation of residues
(Stern and Shull, 1990).

Some residues may undergo evaporation of their H and He. If we assume that
$\breve \gamma$ is the fraction of the generalized formation rate
that will initially produce such residues,
the gas will be replenished by a mass per unit time
$(1-Z)\breve\gamma\Psi$ due the evaporation
of H and He, where $Z$ is the metallicity of the gas and $Z\breve
\gamma\Psi$ is the mass of metals which go into these residues per unit time.
We will assume an instantaneous evaporation.
Based on Tinsley and Cameron (1974) and Van\'ysek (1987), we set
$\breve\gamma\gg\gamma-\breve\gamma$, so
that as a first approximation $\breve\gamma\approx\gamma$.

Adopting the instantaneous recycling
approximation (IRA),  $\tau\sb m\approx 0$. The ejection rate
to the interstellar medium and the rate of mass locked up in
remnants can be simplified as
$$E(t) = R \Psi(t) \eqno(8)$$
where $R$ is the returned fraction to the interstellar gas,
$$R = \int_{m\sb t}^{m\sb u} [m-w_m(m)]\Phi(m) dm \eqno(9)$$
and
$$L(t) = (\varepsilon-R) \Psi(t). \eqno (10)$$
Recalling that $\rho \approx 0$ in equation (7),  that $\breve\gamma \approx
\gamma$, and assuming further that $Z\ll 1$, we can write for the mass rates
$${d\over dt} M\sb g(t) = - (\zeta + \varepsilon - R)\Psi(t) \eqno(11a)$$
$${d\over dt} M\sb s(t) = \zeta\Psi(t) \eqno(11b)$$
$${d\over dt} M\sb r(t) = (\varepsilon - R)\Psi(t) \eqno(11c)$$
Equation (11b) can also be obtained assuming a constant formation rate.
In the framework of the IRA,
it can be easily interpreted: the rate of change of total stellar mass
is only due
to the stars that live forever, which have  masses in the range
$m\sb l<m<m\sb t$.
\bigskip
\bigskip
\centerline{\bf 3. The generalized formation rate and initial mass function}
\bigskip\noindent
The slow rate of growth of the abundances of the heavy
elements  produced by the
metal sink effect due to the refuses (Tinsley and Cameron 1974;
Van\'ysek 1987a,b) is built in our model, and
can be obtained by appropriately chosen fractions
of the generalized formation rate. In order to determine this rate, we will
follow Tinsley and Cameron (1974) and assume that the mass of metals
which go into comets per unit time is at least equal in magnitude to the mass
of
metals which go into the associated star.
Since we have assumed $\breve \gamma\approx\gamma$, it follows that $\gamma
\lapprox\zeta$.
In order to estimate the fractions of the formation
rate, we have used  the stellar
IMF from Miller and Scalo (1979) for $m \geq m\sb l$. We have assumed that the
generalized initial mass function ($\Phi$) for residues is proportional
to $m\sp{-x}$, and re-normalized $\Phi$ in the interval ($m\sb g, m\sb u$)
assuming $\gamma \lapprox \zeta$.
We have taken $m\sb l = 0.1 M\sb\odot$ (Larson, 1992),
$m\sb t = 1 M\sb\odot$ and $m\sb u = 100 M\sb\odot$ (Tinsley, 1980).
It can be shown that if $\gamma\lapprox\zeta$,
then the slope $x$ of the generalized initial
mass function for residues should be $\lapprox 1.8$.
\par
The fractions of the generalized formation rate
can then be computed and we have obtained
$\gamma \approx 0.3$, and $\zeta \approx 0.3$, so that
$\varepsilon \approx 0.4$. The returned fraction
can be computed from equation (9) as $R \approx 0.24$,
where we have taken $w\sb m = 0.7 M\sb\odot$ for
$m \leq 4 M\sb\odot$, and $w\sb m = 1.4 M\sb\odot$ for $m > 4 M\sb\odot$
(Tinsley, 1980).
\bigskip
\bigskip
\vfill\eject
\centerline{\bf 4. Metallicity Distribution}
\bigskip\noindent
Following Tinsley (1980), the metallicity of the gas for the conditions
adopted here can be obtained from the equation
$${d\over dt} Z(t) M\sb g(t) = -Z(1-R)\Psi + y(\varepsilon+\zeta-R)\Psi
\eqno(12)$$
where we have kept the definition of the heavy element yield
$y$ as the total mass of new ejected metals relative to the
mass locked up in stars and remnants,
$$y = {1\over \zeta + \varepsilon - R} \int\sb{m\sb t}\sp{m\sb u} m p\sb{Zm}(m)
\Phi(m)
dm \eqno(13)$$
where $p\sb{Zm}$ is the so-called stellar yield, or the mass fraction
of a star with mass $m$ that is converted to metals and ejected.

{}From equations (11a) and (12) the metallicity can be
integrated as
$$Z(t)= {y(1-\gamma-R)\over \gamma}\left[1-\left({\mu\over\mu_0}\right)
^{{\gamma\over(1-\gamma-R)}}\right] + Z_0\left({\mu\over\mu_0}\right)^{{\gamma
\over(1-\gamma-R)}}. \eqno(14)$$
where $\mu = \mu(t)$, so that $\mu = \mu\sb 0$ and $Z = Z\sb 0$
for $t = 0$.  We set $t=0$ in that instant when the disk reached
its final mass $M$.

In order to show the dependence of the heavy element abundance with
$\gamma$, we will take $y = 0.01$ (Tinsley and
Cameron, 1974; Maciel, 1992) and $R = 0.24$ as discussed in section 3.
We have also adopted an initially unenriched gas, so that $Z\sb 0
\approx 0$.
Figure 1 shows $Z$ as a function
of the ratio $\mu/\mu\sb 0$ for some representative values of $\gamma$.
We see that the smaller is the value for $\gamma$, the greater is
 $Z(\mu\to 0)$. Tentative limits for $\gamma$ are provided, assuming that
$Z(\mu \to 0) \approx Z\sb\odot = 0.02\pm 0.01$ in equation (14).
On the basis of the assumed error, we can see in figure~2 that
values for $\gamma$ in the range 0.19-0.38 are preferred. It can be seen
that the residue mass fraction obtained in section 2 lies approximately
in the middle of this interval.

An analytical expression for the cumulative distribution of stars
of a given metallicity can be derived for the one-zone
model with  metal retention by refuses.
Recalling the definition of the gas and refuse fractionary
superficial densities, equations (2) and (3), respectively,
the fraction of stars born until the metallicity has reached a value $Z$ is
$$S(Z) = {M\sb s\over M\sb{s\sb 1}} = {1-\mu(Z)-\kappa(Z) \over
1-\mu_1-\kappa_1},
\eqno(15)$$
where the subscript 1 indicates present values. From equations (11a) and
(11c) we can write
$$\kappa=\kappa_0 + {\varepsilon-R\over 1-\gamma-R}
(\mu_0-\mu),\eqno(16)$$
where the subscript 0 again indicates initial values.
{}From equations (14), (15) and (16), and recalling the definitions
(2) and (3), we obtain, after some algebra
$$S(Z) = {{a - b\mu_0 {\left[{(Z/Z_1 - 1)-(Z/Z_1 -Z_0/Z_1)(\mu_1/\mu_0)^
{\gamma\over 1-\gamma-R}\over Z_0/Z_1 - 1}\right]}^{{1-\gamma-R\over\gamma}}}
\over a-b\mu_1},\eqno(17)$$
where $a$ and $b$ are constants given by
$$a = 1 - \kappa_0 - {\varepsilon - R\over 1-\gamma-R}\mu_0 \eqno(18)$$
$$b = {\zeta\over 1-\gamma-R}. \eqno(19)$$
Of course, $S$ is normalized, so that form (17) we have $S = 1$
for $Z = Z\sb 1$.

In order to analyze the results of equation (17), we have
varied the initial conditions, namely $\mu_0$, $\kappa_0$
e $Z_0$. We have chosen eight sets of initial conditions,
labeled by letters A to H, which are shown in table 1.
To obtain numerical estimates, we have used the
mass fractions given in section 3, namely $\gamma = 0.3$, $\zeta = 0.3$,
and $\varepsilon = 0.4$. We have further assumed $\mu\sb 1 = 0.1$
(Tinsley, 1980; Pagel and Patchett, 1975), and $R = 0.24$.

Figure 3 shows cases A to D. As a comparison with the
observational data, the asterisks in the figure are
obtained from the differential metallicity distribution of
132 G dwarfs in a cylinder passing through the Sun and perpendicular
to the galactic plane (Pagel, 1989). We have taken $Z\sb 1 = 1.19 Z\sb\odot$,
according to Table 2 of Pagel (1989), which corresponds to the central
value of the largest metallicity bin.

Case A will arise in a disk with non unitary
gas fraction, but with no initial refuses. The distribution will predict even
greater values than the simple model (Schmidt, 1963). This is due to the fact
that when we set $\mu_0\not=1$ and $\kappa_0=0$, we are necessarily accepting
some primordial stars with low metallicities.
Case B is the simple model with metal sink effect.
Cases C and D shows ``prompt initial enrichment'' models in which the burst
of star formation will also lead to the formation of refuses.

Figure 4 shows the effect of increasing $Z_0$ fixing the initial
gas and refuse mass fractions. We can see that the fits to the
observational data are much better, especially for the models with higher
initial heavy element abundances (models G, H). This result is particulary
interesting when we compare models A and E, where the inclusion of
refuses and a very small initial heavy element abundance produce a
large difference in the cumulative distribution at low metallicities.
\bigskip
\bigskip
\centerline{\bf 5. Comets and residues}
\bigskip\noindent
Tinsley (1974) has pointed out that two empirical results provide
powerful constraints on chemical evolution models, namely the G-dwarf
problem and the slow enrichment rate of the ISM.
In our model, the
assumption that comets are like sinks of metals explains easily
this slow enrichment, provided that the slope of the
generalized initial mass function for the residues is
$\lapprox 1.8$. On the other hand, the G-dwarf problem is also
explained if we postulate that a generation of primordial massive stars
will give to the disk an initial metallicity, as well as some initial
remnants. Of course, the initial conditions are connected, so that
the values for $\mu_0$ and $\kappa_0$ are likely to depend on the
value for $Z_0$. The crucial assumption in our model is that
$\breve \gamma\approx\gamma$.

In order to obtain an order-of-magnitude
estimate of the importance
of comets as part of the residues, we have estimated the total
initial residue mass $\gamma \Psi$. Using $0.19<\gamma<0.38$, and
a present value of the generalized formation rate
similar to the star formation rate, $\Psi \approx 10 M\sb\odot$ pc$\sp{-2}$
Gyr$\sp{-1}$ (Tinsley, 1980; Miller and Scalo, 1979),
we have $\gamma\Psi\sim 1.9-3.8 M\sb\odot $pc$\sp{-2}$ Gyr$\sp{-1}$.
The corresponding term for comets can be estimated by
$$Z\breve\gamma\Psi \sim {N\sb c h\sb c M\sb c\over V \tau} \eqno (20)$$
where $N\sb c$ is the number of comets,
$h\sb c$ is the comet galactic scale height, $M\sb c$ is the
average nuclear mass of a comet, $V$ is the total volume considered and
$\tau$ is the system lifetime.
We have first taken into account the whole solar system, where
$N\sb c \sim 2.5 \times 10\sp 6$ (Allen, 1973),
$V \sim 3.1 \times 10\sp{-11}$
pc$\sp 3$ and $h\sb c \sim 4.0 \times 10\sp{-4} $ pc with a radius
of 40 AU (Allen, 1973), $\tau \sim 5$ Gyr, and $M\sb c \sim 10\sp{-16}
M\sb\odot$ (Van\'ysek, 1987), so that $Z\breve\gamma\Psi \sim 6.5 \times
10\sp{-4} M\sb\odot$ pc$\sp{-2}$ Gyr$\sp{-1}$. Taking the average
heavy element abundance during the solar system lifetime $Z \sim 0.01$,
we have $\breve\gamma \Psi \sim 0.06 M\sb\odot$ pc$\sp{-2}$ Gyr$\sp{-1}$,
which is much lower than the $\gamma \Psi$ fraction estimated above.
Assuming now the existence of the so-called
``Massive Oort Cloud'' with $h\sb c \sim 0.10 $ pc
for an adopted radius of $10\sp{4}$ AU (Van\'ysek
1987), $V \sim 4.8 \times 10\sp{-4}$ pc$\sp{3}$,
$\tau \sim 5 $ Gyr, and $M\sb c \sim 10\sp{-16} M\sb\odot$,
we need $N_c \sim 10^{13}$ comets to account for most of
the residue mass, in
agreement with independent results by Stern and Shull (1990),
Van\'ysek (1987a,b) and Greenberg (1974).
\vskip 1.0 true cm
\centerline{\bf Acknowledgements}
\medskip\noindent
H.J.R.-P. wishes to acknowledge gratitude to Dr. L. da Silva
for his support. This work was partially supported by UFRJ,
CAPES, CNPq and FAPESP.
\bigskip
\bigskip
\centerline{\bf References}
\bigskip
\ref Bailey, M.E.: 1988, in M.E. Bailey, D.A. Williams, (eds.), {\it
     Dust in the Universe}, Cambridge University Press, Cambridge, p. 113.
\ref Allen, C.W.: 1973, {\it Astrophysical Quantities}, Athlone.
\ref Ferrini F., Matteucci F., Pardi C., and Penco, U.: 1992, {\it Astrophys.
     J.} {\bf 387}, 138.
\ref Greenberg, J.M.: 1974, {\it Astrophys. J.} {\bf 189}, 19.
\ref Larson, R.B.: 1992, {\it Monthly Notices Roy. Astron. Soc.} {\bf 256},
     641.
\ref Maciel, W.J.: 1992, {\it Astrophys. Space Sci.} {\bf 196}, 23.
\ref Meusinger, H.: 1992, {\it Astrophys. Space Sci.} {\bf 188}, 19.
\ref Miller G.E., and Scalo, J.M.: 1979, {\it Astrophys. J. Suppl.} {\bf 41},
     513.
\ref Pagel, B.E.J.: 1989, in J. E. Beckman and B. E. J. Pagel (eds.),
     {\it Evolutionary Phenomena in Galaxies}, Cambridge University Press,
     Cambridge, p. 201.
\ref Pagel, B.E.J., and Patchett, B.E.: 1975, {\it Monthly Notices Roy.
     Astron. Soc.} {\bf 172}, 13.
\ref Rana, N.C., and Wilkinson, D.A.: 1986, {\it Monthly Notices Roy. Astron.
     Soc.} {\bf 218}, 497.
\ref Schmidt, M.: 1959, {\it Astrophys. J.} {\bf 129} 243.
\ref Schmidt, M.: 1963, {\it Astrophys. J.} {\bf 137} 758.
\ref Stern, S.A, and Shull, J.M.: 1990 {\it Astrophys. J.} {\bf 359}, 506.
\ref Tinsley, B.M.: 1974, {\it Astrophys. J.} {\bf 192}, 629.
\ref Tinsley, B.M.: 1980, {\it Fund. Cosm. Phys.} {\bf 5}, 287.
\ref Tinsley, B.M.: 1981, {\it Astrophys. J.} {\bf 250}, 758.
\ref Tinsley, B.M., and Cameron, A.G.W.: 1974, {\it Astrophys. Space Sci.}
     {\bf 31}, 31.
\ref Truran, J.W., and Cameron, A.G.W.: 1971, {\it Astrophys. Space Sci.}
     {\bf 14}, 179.
\ref Van\'ysek, V.: 1987a, in  {\it Symposium on the diversity and
     similarity of comets}, ESA SP-278, Brussels, p. 745.
\ref Van\'ysek, V.: 1987b, in J. Palou\v s (ed.) {\it 10th IAU European
     Regional Meeting}, Praha, p. 279.
\bye